\documentclass[prl,aps,twocolumn,showpacs]{revtex4}
\usepackage{epsfig}
\usepackage{graphicx}
\usepackage{textcomp}
\newcommand{\Op}[1]{{{\mathrm{\hat{#1}}}}}

\newcommand{\qb}{{\bf q}}

\newcommand{\pb}{{\bf p}}
\newcommand{\xb}{{\bf x}}
\newcommand{\rb}{{\bf r}}
\newcommand{\Xb}{{\bf X}}

\def\p12{p_{12}({\bf q},t)}

\begin{document}
\title{Fast migration and emergent population dynamics}

\author{Michael Khasin$^{1,2}$,  Evgeniy Khain$^2$, and Leonard M. Sander$^1$}
\affiliation{$^1$Department of
Physics, University of Michigan, Ann Arbor, MI 48109-1120, USA}
\affiliation{$^2$Department of Physics, Oakland University, Rochester, MI 48309, USA}



\begin{abstract}
We consider population dynamics on a network of patches, each of which has a the same local dynamics, with different population scales (carrying capacities). It is reasonable to assume that if the patches are coupled by very fast migration the whole system will look like an individual patch with a large effective carrying capacity. This is called a ``well-mixed" system. We show that, in general, it is not true that the well-mixed system has the same dynamics as each local patch. Different global dynamics can emerge from coupling, and usually must be figured out for each individual case. We give a general condition which must be satisfied for well-mixed systems to have the same dynamics as the constituent patches.
\end{abstract}


\maketitle
Many population models consider patches of habitat, each with a local birth-death dynamics,  where the patches are coupled by diffusive migration \cite{Levins, Hanski}. Here we will consider a special case where all the patches have the \emph{same} local dynamics except for a parameter that gives the scale of the local mean population (carrying capacity). That is, we consider a metapopulation which can exchange between patches which are more or less favorable places to be.

Such models are notoriously complex and have rich behavior \cite{Hanski}. However, one particular case seems simple: suppose the migration rate is very fast, faster than any other rate in the problem. In this case all the populations will become equal on average. It is tempting to think that the autonomous dynamics of the total population will be the same as that on each patch with some sort of average carrying capacity, i.e., the whole system will act as one patch carrying the total population \cite{Yaari12}. In this case the system is ``well-mixed''.

In this paper we will show that this common belief is false in general: there are tight restrictions for the type of dynamics that admits this type of homogenization. If these restrictions are not obeyed the dynamics of the emergent behavior of the total population can be different in nature from that of the individual patches. We will derive the restrictions on the dynamics and give examples of the sort of emergent behavior that we have described, first by using mean-field theory and then including fluctuations.

We first formulate in a more precise form the meaning of the idea that a total population of a number of local patches evolves like a larger population on a single patch. Let $X$ be the size of a population. The population dynamics is driven by the birth-death processes $X \rightarrow X+r$, each process associated with the rate $W(X;r)$. We shall assume that the rate can be put into the form
\begin{eqnarray}
\label{eq:rate}
&&W(X;r)=K w(x;r)  \ \  \nonumber \\
&&x=K^{-1} X. \nonumber
\end{eqnarray}
The number $K$ is  the carrying capacity. Now consider a connected network of $N$ patches such that the local dynamics of the population $X_i$ is identical on each patch, but the carrying capacities, $K_i$  are different. That means the local birth-death processes are of the form $X_i\rightarrow X_i+r$, with corresponding rates
\begin{eqnarray}
\label{eq:rate2}
&&W_i(X_i;r)=K_i w(x_i;r),  \ \  \ i=1,...,N,  \\
&&x_i=K_i^{-1} X_i.\nonumber
\end{eqnarray}

The patches are connected by migration. We  define the dynamics of the total population $X=\sum_i X_i$ to be well-mixed if it is driven by transitions $X\rightarrow X+r$, with corresponding rates
\begin{eqnarray}
\label{eq:wmrate}
W(X;r)=\tilde{K} w(x;r) \ \ x=\tilde{K}^{-1} X, \ \  \nonumber
\end{eqnarray}
where $w(x;r)$ are local rates (\ref{eq:rate2}). The number $\tilde{K}$ is the effective carrying capacity of the total population.
This defines  well-mixed population dynamics.

Now assume that the rate of migration between patches is much larger than the local birth-death rates. We  first consider the mean field limit, neglecting fluctuations which is appropriate for $K_i\sim K \gg 1$. The mean-field equations are
\begin{eqnarray}
\label{eq:mf0}
\dot{X}_i&=&\sum_r r W_i(X_i;r)+D\sum_{j\in {\cal I}_i}\left(X_j-X_i\right) \nonumber \\
&=&\sum_r r K_i w(K_i^{-1} X_i;r)+D\sum_{j\in {\cal I}_i}\left(X_j-X_i\right)
\end{eqnarray}
where $D$ is migration rate, which is taken to be identical for all groups, and ${\cal I}_i$ is the set of indices, associated with the patches, connected by migration to patch $i$. We are interested to see if for $D \rightarrow  \infty$, the total population is driven by the homogenized dynamics:
\begin{eqnarray}
\label{eq:mf1}
\dot{X}=\sum_r r \tilde{K} w(\tilde{K}^{-1} X;r),
\end{eqnarray}
for some effective carrying capacity $\tilde{K}$.

Consider a specific example first where the local dynamics displays the  Allee effect \cite{Stephens99}; i.e. in a certain range of parameters the single patch dynamics is bistable: there is one stable state with finite population and another corresponding to local extinction.
Our example for the  local birth-death process is:
\begin{eqnarray}
X_i  &\rightarrow&  0; \quad \ \  \ {\rm Rate} \quad \mu X_i \nonumber \\
2X_i &\rightarrow & 3X_i;  \quad {\rm Rate} \quad (\lambda/2K_i) X_i (X_i-1) \nonumber \\
3X_i &\rightarrow& 2X_i; \quad {\rm Rate} \quad (\sigma/6K_i^2) X_i (X_i-1)(X_i-2), \nonumber
\end{eqnarray}
on each site, where $K_i$ is a local carrying capacity. We note that the rates have the canonical form (\ref{eq:rate2}). The mean-field equations are:
\begin{eqnarray}
\label{eq:mfal}
\dot{X}_i&=& -\mu X_i+\frac{\lambda}{2 K_i}X_i^2-\frac{\sigma}{6 K_i^2}X_i^3 \nonumber \\
&+&D\sum_{j\in {\cal I}_i}\left(X_j-X_i\right)
\end{eqnarray}
For $D=0$ the local stable stationary states are found to be $X_i^* =\{0, \frac{3\lambda K_i}{2\sigma} [1 + (1-\frac{8\sigma\mu}{3\lambda^2} )^{1/2}]\}$, i.e., local bistability (Allee effect) is observed for $\lambda^2 > (8/3)\sigma \mu$.

For large migration rates 
the local populations are approximately equal, $X_i=X/N$, on the time-scale of local dynamics.  The mean-field equation of motion for the total population can be derived
by summing up the equations (\ref{eq:mfal}) to get rid of migration terms, and setting $X_i=X/N$:
\begin{eqnarray}
\label{eq:mfq2}
&&\dot{X}=-\mu X+\frac{\lambda}{2 \tilde{K}}X^2-\frac{\tilde{\sigma}}{6 \tilde{K}^2} X^3  \\
&&\tilde{K}\equiv N^2 \left(\sum_i K_i^{-1}\right)^{-1} \ \ \ \tilde{\sigma}\equiv \frac{\sigma }{N^3}\sum_i \left(\frac{\tilde{K}}{K_i}\right)^2.\nonumber
\end{eqnarray}
Comparing Eqs.(\ref{eq:mfq2}) and (\ref{eq:mfal}) we see that the total population $X$ follows a single-patch dynamics with the \textit{rescaled} rate $\tilde{\sigma}$. Note that for equal  local carrying capacities, $K_i=K_j$ for all $i,j$,   $\tilde{\sigma}= \sigma$. On the other hand, for different carrying capacities it can be shown that $\tilde{\sigma} > \sigma$, i.e., \emph{the well-mixing condition is not satisfied}.

This can lead to a qualitative difference of the local dynamics and dynamics of the total population: specifically, the local dynamics is bistable for $\lambda^2 > \frac{8}{3} \sigma \mu$, while for  $\lambda^2 < \frac{8}{3} \sigma \mu$ the only stable state is extinction of the population. Since $\tilde{\sigma}\ge \sigma$, in a certain range of parameters one can simultaneously have $\lambda^2 >\frac{8}{3} \sigma \mu$ and $\lambda^2 < \frac{8}{3} \tilde{\sigma} \mu$. In this case, even though each local population separately can be bistable, the total population in the presence of fast migration is driven to extinction for any initial distribution of local populations. Next we present an example illustrating this  remarkable dynamics.

\begin{figure}[ht]
\includegraphics[width=9.0cm,clip=]{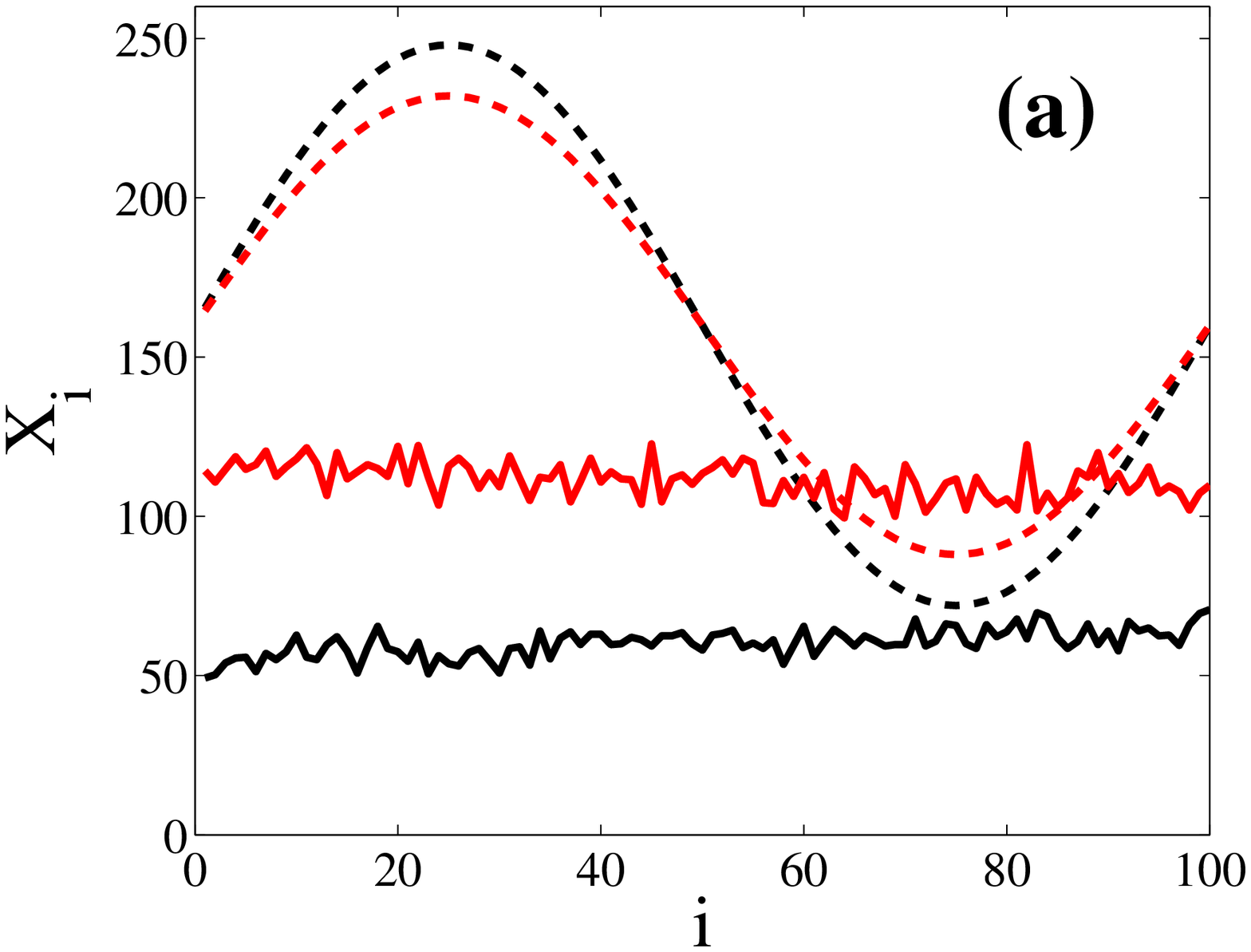}
\includegraphics[width=9.0cm,clip=]{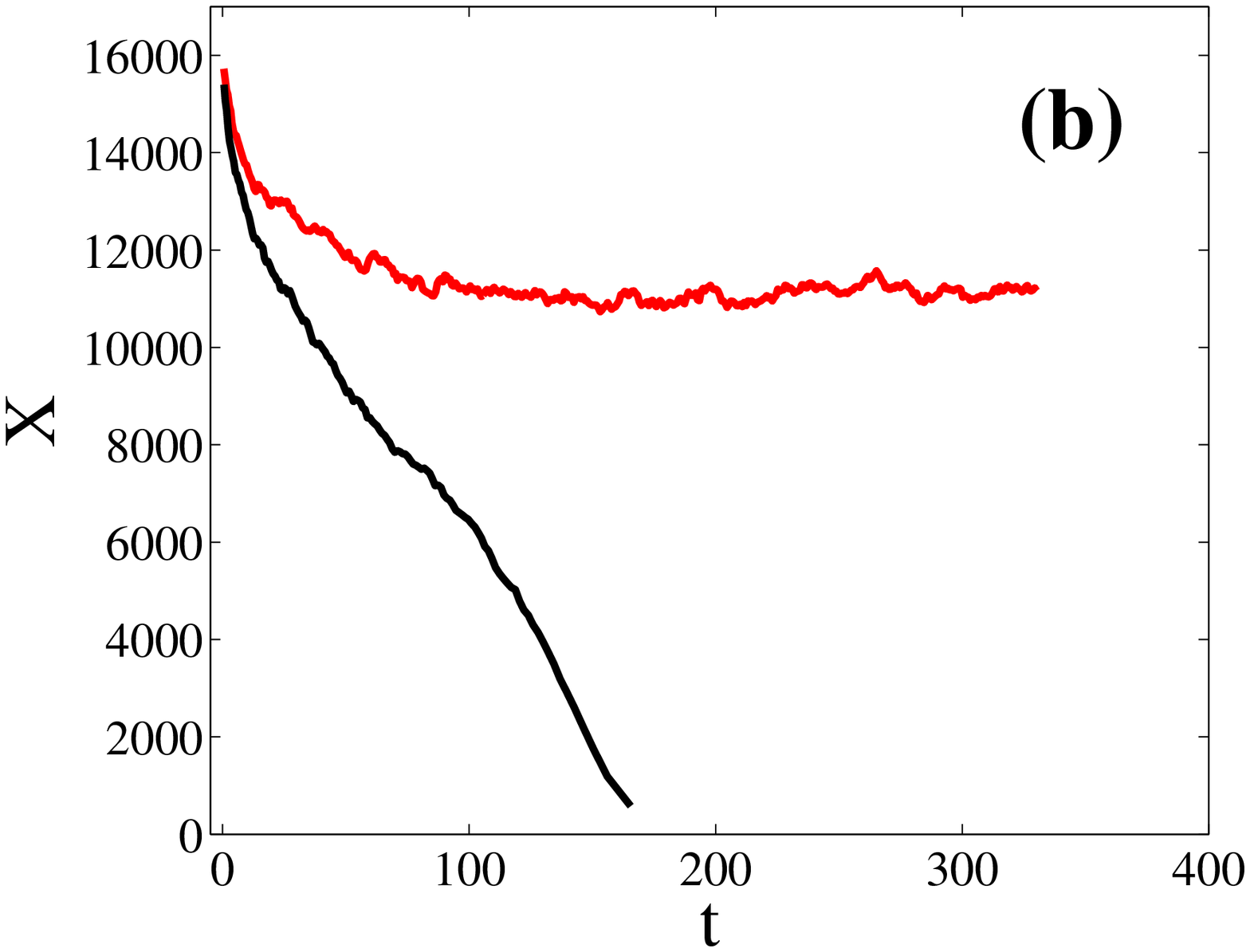}
\caption{(a) Spatial population profiles $X_i$ at intermediate time $t=100$ for two values of normalized variation of carrying capacity: $\epsilon=0.45$ (upper solid line) and $\epsilon=0.55$ (lower solid line). The dashed lines show the corresponding spatial profiles for zero diffusion. (b) The total population $X$ as a function of time for $\epsilon=0.45$ (upper solid line) and $\epsilon=0.55$ (lower solid line). The parameters are: $K=60$, $D=300$,  $\mu=0.09$, $\lambda = 0.2$, $\sigma = 0.15$.} \label{figure1}
\end{figure}

We consider the following  dependence of the local carrying capacity on the (discrete) spatial coordinate $i$:
$$K_i=K(1+\epsilon\sin(2 \pi i/L)),$$
 so that the carrying capacity is a periodic function of $i$, and the parameter $\epsilon$ determines the relative magnitude of these spatial variations, see Fig.~\ref{figure1}a (dashed lines). For sufficiently small $\epsilon$ the nontrivial nonzero state is stable. However, when $\epsilon$ exceeds the threshold determined by $\lambda^2 = \frac{8}{3} \tilde{\sigma}(\epsilon) \mu$, the system is rapidly driven to extinction. Fig.~\ref{figure1}a shows the spatial population profiles $X_i$ for two values of $\epsilon$ (above and below the threshold); Fig.~\ref{figure1}b shows the corresponding time dependence of the total population $X$.

In the general case the necessary condition for the well-mixing is obtained from Eq.(\ref{eq:mf0}) and (\ref{eq:mf1}) following similar manipulations. The condition for well-mixing is:
\begin{eqnarray}
\label{eq:conditionmf}
\sum_{r,i} r K_i w\left( (N K_i)^{-1}X ;r\right)=\sum_r r \tilde{K}  w\left( \tilde{K}^{-1} X ;r\right),\nonumber
\end{eqnarray}
for some $\tilde{K}$.

We can  find the necessary and sufficient condition for well mixing  beyond mean-field theory. Stochastic population dynamics is driven by the master equation for the probability distribution:
\begin{eqnarray}
\label{eq:pop}
\dot{P}\left(\textbf{X}\right)&=&\sum_{i=1}^N \sum_r W_i\left(X_i-r;r \right)P\left(\Xb-r \textbf{e}_i\right) \nonumber \\
&-&\sum_{i=1}^N \sum_r W_i\left(X_i; r\right)P\left(\Xb\right) + D \Op W_0 {P} \left(\textbf{X}\right)
\end{eqnarray}
where $W_i(X_i;r)$ are local transition rates, $(\textbf{e}_i)_j=\delta_{ij}$ and  $D \Op W_0\left(\Xb,\rb\right)$ is the  migration operator:%
\begin{eqnarray}
\label{eq:hop}
W_0\left(\Xb,\rb\right)P(\Xb) &=&\frac{1}{2}\sum_{i=1}^N \sum_{j\in {\cal I}_i} (X_j+1)P(\Xb-\textbf{e}_i+\textbf{e}_j)\nonumber \\
&+& \frac{1}{2}\sum_{i=1}^N \sum_{j\in {\cal I}_i}(X_i+1) P(\Xb+\textbf{e}_i-\textbf{e}_j)\nonumber \\
&-& \frac{1}{2}\sum_{i=1}^N \sum_{j\in {\cal I}_i}(X_i+X_j) P(\Xb)\nonumber
\end{eqnarray}

We use on-site dynamics identical on all patches with different local carrying capacities $K_i$  We also assume the carrying capacities are large, $K_i=K \kappa_i$, $K \gg 1$.
We seek the dynamics of total population $X\equiv \sum_i X_i$ in the limit  $D \rightarrow \infty$.

For $K \gg 1$ we can use the eikonal approximation for the master equation (\ref{eq:pop}) \cite{Kubo73,Gang87,Dykman94,kamenev04,Kessler07, AssafPRE81}, looking for the probability distribution in the form
\begin{eqnarray}
\label{eq:ansatz}
&&{P}\left(\textbf{X},t\right)=e^{-K S\left(\textbf{x},t\right)}  \\
&&\textbf{x}\equiv K^{-1}\textbf{X}. \nonumber
\end{eqnarray}
For the quasistationary distribution
we obtain Hamilton-Jacobi equation for the action $S\left(\textbf{x}\right)$ \cite{llmech}:
\begin{eqnarray}
\label{eq:hj}
&&H\left(\textbf{x},\textbf{p}\right)=0 \nonumber \\
&&p_i=\partial_i S\left(\textbf{x}\right),\nonumber
\end{eqnarray}
associated with the following Hamiltonian:
\begin{eqnarray}
\label{eq:ham}
H\left(\textbf{x},\textbf{p}\right)&=&D H_0\left(\textbf{x},\textbf{p}\right)+ H_1\left(\textbf{x},\textbf{p}\right)\\
H_1\left(\textbf{x},\textbf{p}\right)&=&\sum_{i=1}^N \sum_r \kappa_i w\left(\kappa_i^{-1} x_i ;r\right)\left(e^{p_i r}-1\right)\nonumber \\
H_0\left(\textbf{x},\textbf{p}\right)&=&\frac{1}{2}\sum_{i} \sum_{j\in {\cal I}_i} x_i\left(e^{p_j-p_i}-1\right)\nonumber \\
&+&\frac{1}{2}\sum_{i} \sum_{j\in {\cal I}_i} x_j\left(e^{p_i-p_j}-1\right). \nonumber
\end{eqnarray}
The Hamiltonian (\ref{eq:ham}) generates the fluctuational dynamics of the system, in the sense that the most probable path of the system, leading from $\xb_i$ at $t=t_i$ to $\xb_f$ at $t_f$ can be shown \cite{Dykman94} to correspond to Hamiltonian dynamics generated by (\ref{eq:ham}).  To derive the fluctuational equations of motion for the (rescaled by $K$) total population, we do a canonical transformation from $\xb$ and $\pb$ to $q_i,Q$ and $P_{i},P_Q$, $i=2,3,...,N$:
\begin{eqnarray}
\label{eq:can}
&&Q=\sum_j x_j \ \ \ q_i=x_i \nonumber \\
&&p_1=P_Q \ \ \ p_{i}=P_Q+P_{i}.\nonumber
\end{eqnarray}
In the new variables, the Hamiltonian is in the form $H=D \tilde{H}_0(\{q_i\},Q, \{p_{q_i}\})+\tilde{H}_1(\{q_i\},Q, \{p_{q_i}\},P_Q)$ with $i=2,3,...,N$.

The total population $Q$ and the conjugate momentum $P_Q$ are  slow variables, evolving on the time-scale unity. The variables $q_i,P
_i$ are fast, evolving on the time-scale of migration $\varepsilon \equiv D^{-1}$.
In zero order in $\varepsilon$ the evolution of the total population is driven by the Hamiltonian:
\begin{eqnarray}
\label{eq:hamslow0}
&&H_{slow}\left(Q,P_Q\right)= {H}_1\left( \left\{\tilde{q}_i(Q)\right\},Q, \textbf{0}, P_Q\right), \\
&&\tilde{q}_i(Q)=Q/N, \ \ i=2,3,...,N,\nonumber
\end{eqnarray}
which reflects the adiabatic slaving of the fast local populations $q_i$, by the slow evolution of the total population $Q$, such that instantaneous sizes of the local populations are all equal, due to the fast migration. We outline the derivation below. This derivation generalizes the fast migration result presented in Ref.\cite{KMKS}.

The Hamiltonian equations of motion in new variables and after rescaling of time  $t \rightarrow D t$ are:
\begin{eqnarray}
\label{eq:motion}
\dot{q}_i&=&\partial_{P_i}{H}_0(\qb,Q, \textbf{P})+\varepsilon \partial_{P_i} {H}_1(\qb,Q, \textbf{P},P_Q) \nonumber \\
\dot{P}_{i}&=&-\partial_{q_i}{H}_0(\qb,Q, \textbf{P})-\varepsilon \partial_{x_i} H_1(\qb,Q, \textbf{P},P_Q)  \\
\dot{Q}&=&\varepsilon \partial_{P} H_1(\qb,Q, \textbf{P},P_Q) \nonumber \\
\dot{P_Q}&=&-\partial_Q {H}_0(\qb,Q, \textbf{P})-\varepsilon \partial_Q H_1(\qb,Q, \textbf{P},P_Q),\nonumber
\end{eqnarray}
where we use notation $\qb=(q_2,q_3,...,q_N)$ and $\textbf{P}=(P_2,P_3,...,P_N)$ for brevity.

We look for $q_i=\tilde{q}_i(Q)+\varepsilon q_i^{(1)}$ and $P_i=\varepsilon P_i^{(1)}$, where $\tilde{q}_i(Q)$ is the stationary value of $q_i$ for the instantaneous value of $Q$. In the leading order in $\varepsilon$ the first two lines in equations of motion (\ref{eq:motion}) become
\begin{eqnarray}
\label{eq:motion2}
0&=&\partial_{P_i}{H}_0(\qb,Q, \textbf{P})\nonumber \\
0&=&-\partial_{q_i}{H}_0(\qb,Q, \textbf{P})-\varepsilon \partial_{x_i} H_1(\qb,Q, \textbf{P},P_Q)
\end{eqnarray}
The first of Eqs.(\ref{eq:motion2}) determines $\tilde{q}_i(Q)$, which is $Q/N$ for the Hamiltonian (\ref{eq:ham}). In view of  the functional form of the Hamiltonian, ${H}_0(\qb,Q, \textbf{P})=\textbf{P} \cdot \textbf{h}(\qb,Q)+O(\varepsilon^2)$, we  obtain from the first of Eqs.(\ref{eq:motion2})
\begin{eqnarray}
\label{eq:motion3}
\frac{\partial{H}_0(\qb,Q, \textbf{P})}{\partial q_j}\frac{d \tilde{q}_j}{d Q}+\frac{\partial {H}_0(\qb,Q, \textbf{P})}{\partial Q }=0
\end{eqnarray}
to first order in $\varepsilon$.
Combining the second of Eqs.(\ref{eq:motion2}) and Eq.(\ref{eq:motion3}) we derive
\begin{eqnarray}
\label{eq:motion5}
-\varepsilon \frac{\partial H_1(\qb,Q, \textbf{P},P_Q)}{\partial{q_j}}\frac{d \tilde{q}_j}{d Q}=-\frac{\partial {H}_0(\qb,Q, \textbf{P})}{\partial Q}
\end{eqnarray}
Using Eqs.(\ref{eq:motion5}) and coming back to the original dimensional time we rewrite the last two Eqs.(\ref{eq:motion})  in the form
\begin{eqnarray}
\label{eq:motion6}
\dot{Q}&=& \partial_{P} H_1(\tilde{\qb}(Q),Q, \textbf{0},P_Q)=\partial_{P} H_{slow}(Q,P) \nonumber \\
\dot{P}&=&- \partial_{x_j} H_1(\tilde{\qb}(Q),Q, \textbf{0},P_Q)\frac{d \tilde{q}_j}{d Q}\nonumber \\
&-& \partial_Q H_1(\tilde{\qb}(Q),Q, \textbf{0},P_Q)=-\partial_Q H_{slow}(Q,P),\nonumber
\end{eqnarray}
where $H_{slow}(Q,P_Q)\equiv  H_1(\tilde{\qb}(Q),Q, \textbf{0},P_Q)$, Eq.(\ref{eq:hamslow0}).

For simplicity of notation we omit the subscript of $P_Q$ in what follows. Using Eqs.(\ref{eq:ham}) and (\ref{eq:hamslow0}) we derive:
\begin{eqnarray}
\label{eq:hamslow1}
&&H_{slow}\left(Q,P\right)= \sum_r \sum_{i=1}^N \kappa_i w\left( \frac{Q}{N\kappa_i} ;r\right)\left(e^{P r}-1\right).
\end{eqnarray}
This Hamiltonian drives the fluctuational dynamics of the total population $Q$.

In order that the total population follows an effective single-site dynamics with a rescaled carrying capacity (well-mixing), the Hamiltonian (\ref{eq:hamslow1}) must have the form of a single-site Hamiltonian:
\begin{eqnarray}
\label{eq:hamslow2}
&&H\left(Q,P\right)= \sum_r  \tilde{\kappa} w\left( \tilde{\kappa}^{-1}Q ;r\right)\left(e^{P r}-1\right).
\end{eqnarray}
The equivalence of expressions (\ref{eq:hamslow1}) and (\ref{eq:hamslow2}) gives the necessary and sufficient condition for well-mixing:
\begin{eqnarray}
\label{eq:condition}
\sum_{i=1}^N \kappa_i w\left( \frac{Q}{N\kappa_i} ;r\right)=\tilde{\kappa} w\left( \tilde{\kappa}^{-1}Q ;r\right), \ \ \ \forall r,
\end{eqnarray}
where $\tilde{\kappa}\equiv \tilde{K} K^{-1}$ is the rescaled effective carrying capacity.
To make further progress we assume that the on-site birth-death rates $w\left( x ;r\right)$ can be Taylor-expanded around $x=0$:
\begin{eqnarray}
\label{eq:taylor}
w\left(x ;r\right)=\sum_{n=0}^{\infty} a_n^r x^n, \ \ \ \forall r, x.\nonumber
\end{eqnarray}
Then condition (\ref{eq:condition}) is equivalent to
\begin{eqnarray}
\label{eq:condition2}
a^r_n\left[\tilde{\kappa}_i^{1-n}-N^{-n}\sum_{i=1}^N \kappa_i^{1-n}\right]=0, \ \ \ \forall r, n.
\end{eqnarray}
The general solution of this equation for varying carrying capacities, $\kappa_i \neq \kappa_j$ for some $i,j$, is
\begin{eqnarray}
\label{eq:condition3}
&&a^r_n=a_1^r\delta_{n,1}+a_{n^*}^r \delta_{n,n^*}  \ \ \ \forall r \nonumber \\
&&\tilde{\kappa}=N \left[N^{-1}\sum_{i=1}^N \kappa_i^{1-n^*}\right]^{\frac{1}{1-n^*}},\nonumber
\end{eqnarray}
where $n^*$ is a fixed natural number. In other words, a rescaled carrying capacity can be found if and only if the on-site birth-death rates $w\left( x ;r\right)$ have the form
\begin{eqnarray}
\label{eq:condition4}
w\left( x ;r\right)=a_1^r x+a_{n^*}^r x^{n^*}  \ \ \ \forall r.
\end{eqnarray}
We note that for identical carrying capacities, $\kappa_i =1$ for all $i$, the condition (\ref{eq:condition2}) is satisfied for arbitrary $a_n^r$ and $\tilde{\kappa}=N$, i.e., for identical carrying capacity the total population evolves like a single-site population with a carrying capacity rescaled by the number of sites $N$.

Many population dynamics models have the form (\ref{eq:condition4}) for $n^*=0,2$. For $n^*=0$, as in Poissonian process \cite{Kampen},
\begin{eqnarray}
\label{eq:capacity0}
\tilde{\kappa}=N\left(N^{-1}\sum_{i=1}^N \kappa_i\right),\nonumber
\end{eqnarray}
i.e., the carrying capacity is $N$ times the arithmetic mean of the local carrying capacities. For $n^*=2$, as in logistic growth \cite{Balakrishnan},
\begin{eqnarray}
\label{eq:capacity2}
\tilde{\kappa}=N\left(N^{-1}\sum_{i=1}^N \kappa_i^{-1}\right)^{-1},\nonumber
\end{eqnarray}
i.e., the carrying capacity is $N$ times the harmonic mean of the local carrying capacities.

%

In this paper we have shown that a notion that seems to be completely obvious is false. It is not true, in general, that if we mix very fast, a metapopulation with identical dynamics on patches of habitat (with different carrying capacities) will synchronize to act like a single population with the \emph{same} dynamics.

The notion is true in very simple special cases: in effect, if the local dynamics is characterized by just one parameter (the carrying capacity) then fast migration can only average things. Most examples that come to mind are exactly of this sort. But if, as in the case of the Allee effect, there is another relevant parameter, then it can change too, and the qualitative behavior of the synchronized population can be \emph{different from that of any individual patch.}

For the general case, only if Eq. (\ref{eq:condition}) is obeyed, do we have the same dynamics in the well-mixed case as for the individual patches. If the rates are polynomials in the populations, only if we have linear plus one other power, do we conserve the dynamics.

We thank B. Meerson for many discussions and interest in this work. M.K. thanks M. Dykman for discussions on the adiabatic approximation. M.K. is grateful to the University of Michigan for hospitality.


\end{document}